# Level Statistics of Stable and Radioactive Nuclei

M. A. Jafarizadeh[a,b], N. Fouladi[c], H. Sabri[c*], B. Rashidian Maleki[c]

[a] Department of Theoretical Physics and Astrophysics, University of Tabriz, Tabriz 51664, Iran.

[b] Research Institute for Fundamental Sciences, Tabriz 51664, Iran.

[c] Department of Nuclear Physics, University of Tabriz, Tabriz 51664, Iran.

[*] h-sabri@tabrizu.ac.ir

**Abstract**

The spectral statistics of nuclei undergo through the major forms of radioactive decays ($\alpha$, $\beta^-$ and $\beta^+$ (or $EC$)) and also stable nuclei are investigated. With employing the MLE technique in the nearest neighbor spacing framework, the chaoticity parameters are estimated for sequences prepared by all the available empirical data. The ML-based estimated values propose a deviation to more regular dynamics in sequences constructed by stable nuclei in compare to unstable ones. In the same mass regions, nuclei transmitted through $\alpha$ decay explore less regularity in their spectra in compare to other radioactive nuclei.

**Introduction**

The investigations of spectral statistics and non-linear dynamics in different nuclei have been regarded as the interesting topics in recent years. In common analyses [1-2], the fluctuation properties of quantum system's spectra are compared with the prediction of Random Matrix Theory (RMT). This model describes a chaotic system by an ensemble of random matrices subject only to the symmetry restrictions [3-7]. Systems with time reversal symmetry such as atomic nuclei are described by Gaussian Orthogonal Ensemble (GOE). On the other hand, systems whose classical dynamics are everywhere rigorous in the phase space, are well characterized by Poisson distribution.

Different statistics such as Nearest Neighbor Spacing Distribution (NNSD) and *etc* [8-11] have been employed to describe the statistical situation of considered systems in related to these limits. In NNSD framework (as the most commonly used statistics), a Least Square Fit (LSF) has been carried out to compare the spacing of each sequences with some well-known distribution such as Abul-Magd and *etc* [12-15]. The value of every distribution's parameter(s) describes the deviation to regular or chaotic dynamics while the LSF technique, proposes some unusual uncertainties and also a deviation to more chaotic dynamics [16]. To avoid these disadvantages in estimation processes, Maximum Likelihood Estimation (MLE) technique have employed [16] which yields very exact results with low uncertainties in compare to LSF-based estimated results (estimated values yield accuracies which is closer to Cramer-Rao Lower Bound (CRLB)).

Recently, different statistical analyses have been accomplished on nuclear system's spectra to obtain the statistically relevant samples. The spectral statistics of nuclei classified in different mass regions have been described by shriner *et al* [2] and the investigation of fluctuation properties of nuclei classified as their excitation energies ratios [17-19] which have carried by Abul-Magd *et al*, are some of these comprehensive analyses. The results of such description suggest an apparent relation between the chaocity of considered systems and some parameters such as mass, deformation parameter and *etc*. In the present study, we investigate the statistical properties of different nuclei classified according to their decay modes to describe the relation between chaocity and stability in different mass regions. Therefore, we consider four categories, nuclei undergo through $\beta^-$ emission, nuclei exhibit $\beta^+$ (or *EC*) decay, nuclei transmitted through $\alpha$ decay and also stable nuclei.

To describe the spectral fluctuations of considered categories in different mass regions, sequences constructed by nuclei in which the spin-parity $J^\pi$ assignments of at least five consecutive levels are definite. For the relative abundances of $2^+$ and $(1/2)^+$ levels, respectively in the even and odd-mass nuclei, we have used all the available empirical data [20-21] for only these levels in selected nuclei. The ML-based estimated values for chaoticity parameter (Abul-Magd distribution's parameter) propose more regular dynamics for sequences of stable nuclei in compare to radioactive ones. Also, in all considered categories, i.e. stable and three decay modes, heaviest nuclei explore more regular dynamics in compare to lightest nuclei. The ML-based estimated values suggest a deviation to regular dynamic for even-mass nuclei in compare to odd-mass ones in similar mass regions of all stable and radioactive nuclei, too.

This paper is organized as follows, section 2 dealt with reviewing a statistical approach and Maximum Likelihood Estimation (MLE) method. Data sets which have used in this analysis introduced in section 3 and finally, section 4 contains the numerical results obtained by applying the MLE to different sequences. Section 5 is devoted to summarize and some conclusion based on the results given in section 4.

## 2. The method of Statistical analysis

The spectral fluctuations of low-lying nuclear levels have been considered by different statistics such as Nearest Neighbor Spacing Distribution (NNSD) [2-3], linear coefficients between adjacent spacing [2] and Dyson-Mehta $\Delta_3(L)$ [9-11] which based on the comparison of statistical properties of nuclear spectra with the predictions of Random Matrix Theory (RMT). The NNSD, or $P(s)$ functions, is the observable most commonly used to analyze the short-range fluctuation properties in the nuclear spectra. The NNSD statistics requires to complete (few or no missing levels) and pure (few or no unknown spin-parities) level scheme [1-2] where these condition are available for a limited number of nuclei. Therefore, we in need to combine different level schemes to construct sequences. To compare the different sequences to each other, each set of energy levels must be converted to a set of normalized spacing, namely, each sequence must be unfolded. The unfolding process has been described in detail in Refs.[2]. Here, we briefly outline the basic ansatz and summarize the results. To unfold our spectrum, we had to use some levels with same symmetry. This requirement is equivalent with the use of levels with same total quantum number ($J$) and same parity. For a given spectrum $\{E_i\}$, it is necessary to separate it into the fluctuation part and the smoothed average part, whose behavior is nonuniversal and cannot be described by RMT [1]. To do so, we count the number of the levels below $E$ and write it as

$$N(E) = N_{av}(E) + N_{fluct}(E) \qquad , \qquad (2.1)$$

Then we fix the $N_{av}(E_i)$ semiclasically by taking a smooth polynomial function of degree 6 to fit the staircase function $N(E)$. We obtain finally, the unfolded spectrum with the mapping

$$\{\tilde{E}_i\} = N(E_i) \qquad , \qquad (2.2)$$

This unfolded level sequence $\{\tilde{E}_i\}$ is obviously dimensionless and has a constant average spacing of 1, but the actual spacing exhibits frequently strong fluctuation. The nearest neighbor level spacing is defined as $s_i = (\tilde{E}_{i+1}) - (\tilde{E}_i)$. Distribution $P(s)$ will be in such a way in which $P(s)ds$ is the probability for the $s_i$ to lie within the infinitesimal interval $[s, s+ds]$. For nuclear systems with time reversal symmetry which spectral spacing follows Gaussian Orthogonal Ensemble (GOE) statistics, the NNS probability distribution function is well approximated by Wigner distribution [1]

$$P(s) = \frac{1}{2}\pi s e^{-\frac{\pi s^2}{4}} \qquad , \qquad (2.3)$$

While exhibits the chaotic properties of spectra. On the other hand, the NNSD of systems with regular dynamics is generically represented by Poisson distribution

$$P(s) = e^{-s} \qquad , \qquad (2.4)$$

Statistical investigations accomplished on nuclear system's spectra propose intermediate situations of considered systems between these limits. To compare the spectral statistics with regular and chaotic limits quantitatively, different distribution functions have been proposed [12-15]. One of popular distribution is Abul-Magd distribution [12] which was derived by assuming that, the energy level spectrum is a product of the superposition of independent subspectra, which are contributed respectively from localized eigenfunctions onto invariant (disjoint) phase space. This distribution is based on the Rosenzweig and Porter random matrix model [15]. The exact form of this model is complicated and its simpler form is proposed by Abul-Magd *et al* in Ref.[12] as:

$$P(s,q) = [1 - q + q(0.7 + 0.3q)\frac{\pi s}{2}] \times \exp(-(1-q)s - q(0.7+0.3q)\frac{\pi s^2}{4}) \quad , \quad (2.5)$$

Where interpolates between Poisson ($q=0$) and Wigner ($q=1$) distributions. In common considerations, a least square fit (LSF) carried out to compare the Abul-Magd distribution with sequences while the value of distribution's parameter describes the chaotic or regular dynamics. The LSF-based estimated values have some unusual uncertainties and also exhibit more deviation to chaotic dynamics. Consequently, it is almost impossible to perform any reliable statistical analysis in some sequences. Recently [16], we have employed the Maximum Likelihood Estimation (MLE) technique to estimate the parameter of distributions with more precision, i.e. estimated values yield accuracies which are closer to Cramer-Rao Lowe Bound (CRLB). Also, this technique yields results which are almost exact in all sequences, even in cases with small sample sizes, where other estimation methods wouldn't achieve the appropriate results. The MLE estimation procedure has been described in detail in Ref.[16]. Here, we outline the basic ansatz and summarize the results.

- **The ML-based results for Abul-Magd distribution**

The MLE method provides an opportunity for estimating exact result with minimum variation. In order to estimate the parameter of distribution, Likelihood function is considered as product of all $P(s)$ functions,

$$L(q) = \prod_{i=1}^{n} P(s_i) = \prod_{i=1}^{n} [1 - q + q(0.7 + 0.3q)\frac{\pi s_i}{2}] e^{-(1-q)s_i - q(0.7+0.3q)\frac{\pi s_i^2}{4}} \quad , \quad (2.6)$$

Then, with taking the derivative of the log of likelihood function (2.6) respect to its parameter ($q$) and set it to zero, i.e., maximizing the likelihood function, the following relation for desired estimator (see Appendix (C) of Ref.[16] for more details) is obtained

$$f : \sum \frac{-1 + (0.7 + 0.6q)\frac{\pi s_i}{2}}{[1 - q + q(0.7 + 0.3q)\frac{\pi s_i}{2}]} + \sum s_i + (0.7 + 0.6q)\frac{\pi s_i^2}{4} \quad , \quad (2.7)$$

We can estimate "$q$" by high accuracy via solving above equation by Newton-Raphson method which is terminated to the following result,

$$q_{new} = q_{old} - \frac{F(q_{old})}{F'(q_{old})} =$$

$$= q_{old} - \frac{\sum \frac{-1 + (0.7 + 0.6q_{old})\frac{\pi s_i}{2}}{[1 - q_{old} + q_{old}(0.7 + 0.3q_{old})\frac{\pi s_i}{2}]} + \sum s_i - (0.7 + 0.6q_{old})\frac{\pi s_i^2}{4}}{\sum \frac{[0.3\pi s_i][1 - q_{old} + q_{old}(0.7 + 0.3q_{old})\frac{\pi s_i}{2}] - [-1 + (0.7 + 0.6q_{old})\frac{\pi s_i}{2}]^2}{[1 - q_{old} + q_{old}(0.7 + 0.3q_{old})\frac{\pi s_i}{2}]^2} - \sum 0.15\pi s_i^2} \quad , \quad (2.8)$$

Also we have used the difference of both sides of equation (2.8) to obtain the decreasing of uncertainty for estimated values, namely the CRLB for Abul-Magd distribution is defined as [16]

$$Var(\hat{q}) \geq \frac{1}{MF(q)}$$

Where $M$ represents the number of samples and $F(q)$ is used to describe the Fisher information.

## 3. Data sets of stable and radioactive nuclei

A discussion of different decay modes in nuclei caused to the topic of nuclear stability [22-23]. The vast majority of nuclei found in the earth are stable while radioisotopes are unstable ones. In an attempt to reach a more stable arrangement of its protons and neutrons, the nucleus will spontaneously decompose to form a different nucleus while during this procedure, giving off radiation in the form of atomic particles or high energy rays. This decay occurs at a constant, predictable rate that is referred to as half-life. The major forms of radioactive decays are nuclear beta decays included $\beta^-$ and $\beta^+$ (or $EC$) decays and also $\alpha$ decay modes. On the other hand, stable nuclei wouldn't undergo through these kinds of decays and consequently, are non-radioactive. The stable nuclei and also nuclei undergoes through different decay modes are displayed in Figure1.

To analyze the spectral statistics in different mass regions and also to describe the effect of pairing on fluctuation properties of energy spectra for these four categories, i.e. stable nuclei and three categories of radioactive nuclei, we have considered 31 sequences with at least 25 spacing as have introduced in Tables 1-2. In order to prepare sequences by different nuclei with the available empirical data taken from Refs.[20-21], we have followed the same method given in Ref.[2]. Namely, we consider nuclei in which the spin-parity $J^\pi$ assignments of at least five consecutive levels are definite. In cases where the spin-parity assignments are uncertain and where the most probable value appeared in brackets, we admit this value. We terminate the sequence in each nucleus when we reach at a level with unassigned $J^\pi$. We focus on the $2^+$ levels for even mass and $(1/2)^+$ levels for odd-mass nuclei for their relative abundances in specified nuclei. Other levels, i.e. $4^+$ and $6^+$ levels for even-mass nuclei or $(3/2)^+$ and $(5/2)^+$ levels for odd-mass ones, wouldn't consider in this analysis while we couldn't find enough samples for different categories and mass regions by the available empirical data. In this approach, we achieved 165 nuclei of stable ones, i.e. 96 even-mass in addition to 69 odd-mass nuclei such as $^{10}B$, $^{43}Ca$, $^{70}Ge$, $^{94}Zr$, $^{135}Ba$, $^{174}Yb$, $^{194}Pt$ and *etc*, 22 even-mass nuclei undergo through $\alpha$-particle emission, for example $^{144}Nd$, $^{154}Dy$, $^{174}Hf$, $^{234}U$, $^{242}Pu$ and *etc*, 130 nuclei, i.e. 69 even-mass and 61 odd-mass nuclei, undergo through $\beta^-$ decays, for instance as $^{24}Na$, $^{63}Ni$, $^{80}Br$, $^{101}Mo$, $^{128}Te$, $^{160}Gd$, $^{239}U$ and *etc* and 165 nuclei, i.e. 104 even-mass and 61 odd-mass nuclei which exhibit $\beta^+$ (or $EC$) decay such as $^{21}Na$, $^{48}V$, $^{71}As$, $^{118}Xe$, $^{178}W$, $^{188}Pt$ and *etc*.

## 4. Numerical results

In the present analysis, we examine the effect of stability and also different decay modes as probe of statistical properties of nuclear structure. With regard to complete theoretical studies and also experimental evidences of the major forms of radioactive decays[20-23], we tend to classify nuclei in different mass regions of these four categories, i) stable nuclei, ii) nuclei undergoes through $\alpha$ decay, iii) nuclei exhibits $\beta^-$ decay and iv) nuclei suffers during $\beta^+$ (or $EC$ )decay mode.

Although nuclei decay by multiple alternatives energetically possible decay modes, we wouldn't consider all of them in this analysis since the available empirical data wouldn't provide pure and complete sequences in all considered categories and mass regions. Also, we have considered the dominant decay mode for considered nuclei in the present study.

Since, the investigation of the majority of short sequences yields an overestimation about the degree of chaoticity measured by the "q" (Abul-Magd distribution's parameter)[12], therefore, we wouldn't concentrate only on the implicit value of "q" and examine a comparison between the amounts of "q" in the same mass regions for different categories. Therefore, the smallest values of "q" explain more regular dynamics and vice versa.

The considered sequences are unfolded and analyzed with the help of MLE technique, namely, the ML–based estimated value for chaoticity parameter ("q" of Abul-Magd distribution) yield as the converging values of iterations (2.8), where as an initial values we have chosen the values of parameters obtained by LSF method.

A comparison of the chaocity degrees for different even-mass sequences described by "q", presented in Table1. As have explained in previous section, we have employed only $2^+$ levels for even-mass nuclei and also, we have considered sequences with at least 25 samples in this analysis.

| Mass region | stable nuclei | | $\beta^-$ category | | $\beta^+$ category | | $\alpha$ category | |
|---|---|---|---|---|---|---|---|---|
| | N | q | N | q | N | q | N | q |
| $A \leq 50$ | 160 | $0.52 \pm 0.08$ | 137 | $0.64 \pm 0.12$ | 229 | $0.72 \pm 0.10$ | - | |
| $50 < A \leq 100$ | 447 | $0.59 \pm 0.11$ | 250 | $0.71 \pm 0.09$ | 227 | $0.81 \pm 0.13$ | - | |
| $100 < A \leq 150$ | 57 | $0.41 \pm 0.06$ | 187 | $0.43 \pm 0.08$ | 304 | $0.52 \pm 0.09$ | 99 | $0.61 \pm 0.07$ |
| $150 < A \leq 180$ | 258 | $0.32 \pm 0.07$ | 25 | $0.37 \pm 0.06$ | 120 | $0.39 \pm 0.03$ | 43 | $0.48 \pm 0.10$ |
| $180 < A \leq 210$ | 337 | $0.36 \pm 0.04$ | - | | 70 | $0.43 \pm 0.11$ | 30 | $0.53 \pm 0.05$ |
| $A \geq 230$ | - | | - | | - | | 122 | $0.35 \pm 0.08$ |

Table1. The ML-based estimated values for "q" (Abul-Magd distribution's parameter) in the sequences prepared by even-mass nuclei. *N* denotes the number of samples in each sequences.

The chaoticity parameter, i.e. Abul-Magd distribution's parameter, which exhibits the spectral statistics of odd-mass nuclei given in Table2. Considered sequences for stable and radioactive odd-mass nuclei undergo through $\beta^-$ and $\beta^+$ (or $EC$) modes prepared by only $(1/2)^+$ levels for their relative abundance in these nuclei. Also, we couldn't find odd-mass nuclei transmitted through $\alpha$ – decay with at least 5 consecutive $(1/2)^+$ levels and therefore, this category wouldn't appear in this table. A summary of the chaocity degrees of these considered sequences, namely the "q" values for these 31 sequences, presented in Figure2.

| Mass region | stable nuclei | | $\beta^-$ category | | $\beta^+$ category | |
|---|---|---|---|---|---|---|
| | N | q | N | q | N | q |
| $A \leq 50$ | 70 | $0.55 \pm 0.09$ | 34 | $0.69 \pm 0.11$ | 96 | $0.75 \pm 0.14$ |
| $50 < A \leq 100$ | 117 | $0.62 \pm 0.08$ | 160 | $0.72 \pm 0.07$ | 151 | $0.83 \pm 0.11$ |
| $100 < A \leq 150$ | 536 | $0.43 \pm 0.05$ | 192 | $0.45 \pm 0.06$ | 196 | $0.54 \pm 0.10$ |
| $150 < A \leq 180$ | 70 | $0.35 \pm 0.11$ | 59 | $0.38 \pm 0.08$ | - | |
| $180 < A \leq 210$ | 30 | $0.39 \pm 0.07$ | - | | - | |
| $A \geq 230$ | - | | 46 | $0.31 \pm 0.06$ | - | |

Table2. The ML-based estimated values of "q" (Abul-Magd distribution's parameter) in the sequences prepared by odd-mass nuclei. $N$ denotes the number of samples in each sequences.

We plot a relation between the chaoticity parameter and the numbers of $Z$ and $N$ for different nuclei classified in these four categories which presented in Figures (3-7). We wouldn't present the LSF- based estimated results while similar to what have persisted in Ref.[16], exhibit same tendency in the different sequences but with less regular dynamics and also great uncertainty in compare to ML-based estimated values. For instance, the LSF-based estimated value for sequence of odd-mass nuclei undergo through $\beta^+$ -decay mode in the $50 < A \leq 100$ mass region is $q = 1.09 \pm 0.42$. As already mentioned in previous sections, due to the presence of noticeable uncertainty in the LSF estimated values (because of high level variance of estimators), it is almost impossible to do any reliable statistical analysis of some sequences.

From these tables and figures, we see the more regularity of even-mass nuclei in compare to odd-mass ones in all mass regions, similar to the predictions of Refs.[25-26]. It means, the pairing force which is known as the regular part of nuclear force, is weaker between the single particle and collective degrees of freedom in odd-mass nuclei than even-mass ones. Also, similar to the predictions of Refs.[2,6], the lightest nuclei in all categories explore statistical behavior very close to GOE limits while the heaviest ones display a deviation to more regular dynamics.

The ML-based estimated values for chaocity degree propose a deviation to more regularity by stable nuclei in compare to nuclei undergo through these three major forms of radioactive decays. It means, in the same mass regions, stable nuclei have more regular spectra in compare to unstable (radioactive) ones. To investigate the relation between the chaocity degrees of considered sequences by quadrupole deformation value and describe the effect of stability (or different decay modes) on it, we have determined the calculated mean deformation parameter as [15]

$$\langle \beta_2 \rangle = \sum_k N_k \beta_k^2 \Big/ \sum_k N_k$$

Where $N_k$ represents the number of levels of nucleus $k$ which have been involved in the analysis and $\beta_k^2$ is the quadrupole deformation parameter of nucleus $k$ taken from the Ref.[24]. The calculated mean deformation values, $\langle \beta_2 \rangle$, for considered sequences presented in Figure8. A comparison between Figures 2 and 8, suggests an obvious relation between chaocity and deformation, i.e. the chaoticity degrees of sequences decrease with increasing of $\beta$.

Considering the estimated "q" values given in Tables 1-2 and relation between chaocity and deformation of different sequences, we can deduce the following important facts:

i) The ML-based estimated results listed in Tables (1-2) propose more chaotic dynamics for sequences of nuclei emit $\alpha$ – particles in compare to other radioactive nuclei. Since, these nuclei are spherical (magic or semi magic) ones which are expected to have shell model spectra, these results confirm the prediction of GOE [2,7].

ii) The ML-based estimated values for chaocitiy parameter suggest a deviation to more regular dynamic (Poissson limit) for different sequences of $\beta^-$ category. The majority of nuclei undergoing through $\beta^-$-decay are deformed ones where the more regular dynamics for them may be interpreted as AbulMagd-Weidenmuller chaoticity effect. As have proposed in Ref.[27] by Paar *et al*, the identity of nucleons make impossible to define the rotation for spherical nuclei and therefore, rotation of nuclei contribute to the suppression of their chaotic dynamics. It means, the spherical nuclei explore more chaotic dynamics in compare to deformed ones. The same results, i.e. deviation to regularity suggested by nuclei undergo through $\beta^-$, may be achieved by probability distribution of states. $\beta^-$-decay satisfies the maximum randomness condition because after decay, a beta particle and an anti neutrino is given out, so the number of particles, therefore possible micro states increases. The probability of microstates describe by hypergeometric distribution where if the number of these states increase, the hypergeometric distribution is approximated by Poisson distribution.

## 5. Summary

In the present paper, we study the spectral statistics of radioactive nuclei classified according to their major decay mode and also stable ones in the nearest neighbor spacing statistics framework. With using all the available experimental energy levels, sequences prepared and then, the MLE technique is employed to estimate the chaoticity parameter with more precision. The difference in the chaoticity parameter of each categories and also mass regions is statistically significant. Our analysis shows that the chaoticity parameter is smaller in the stable nuclei than the unstable ones undergo through decay modes. Also, spherical nuclei undergo through $\alpha$ – particle emission exhibit less regular dynamics in related to other decay modes. On the other hand, nuclei undergo through $\beta^-$-decay exhibit more regular dynamics in compare to other major decay modes. The apparent regularity of the spectrum is because of the approximate conservation of the quantum number describes the collective degrees of freedom. Our results may be interpreted that the pairing force between the single particle and collective degrees of freedom is weaker in odd-mass nuclei than even-mass nuclei.

# Figure caption

Figure1(color online). A description of nuclei undergo through different radioactive decay modes and also stable nuclei, taken from Ref.[20].

Figure2(color online). The ML-based "q" values for considered sequences. In each mass region, stable nuclei explore more regular dynamics in compare to other categories.

Figure3(color online). Chaoticity parameter (Abul-Magd distribution's parameter) versus the number of protons (Z) and neutron (N) for stable nuclei.

Figure4(color online). Chaoticity parameter (Abul-Magd distribution's parameter) versus the number of protons (Z) and neutron (N) for nuclei undergo through $\beta^- -$ decay.

Figure5(color online). Similar to Figure.3, for nuclei undergo through $\beta^+ -$ decay.

Figure6(color online). Similar to Figure.3, for nuclei emit $\alpha -$ particle.

Figure7(color online). Chaoticity parameter (Abul-Magd distribution's parameter) versus the number of protons (Z) and neutron (N) for all nuclei considered in this analysis.

Figure8(color online). The calculated mean deformation values $\langle \beta_2 \rangle$ for considered sequences.

Figure1.

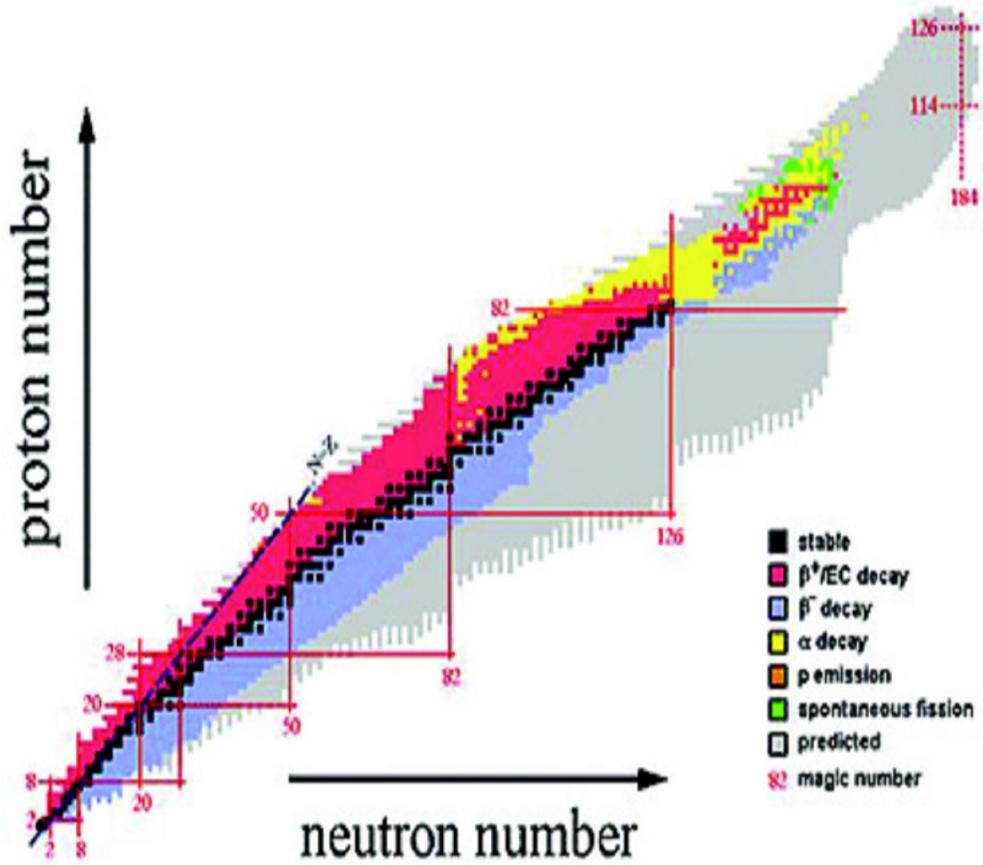

Figure2.

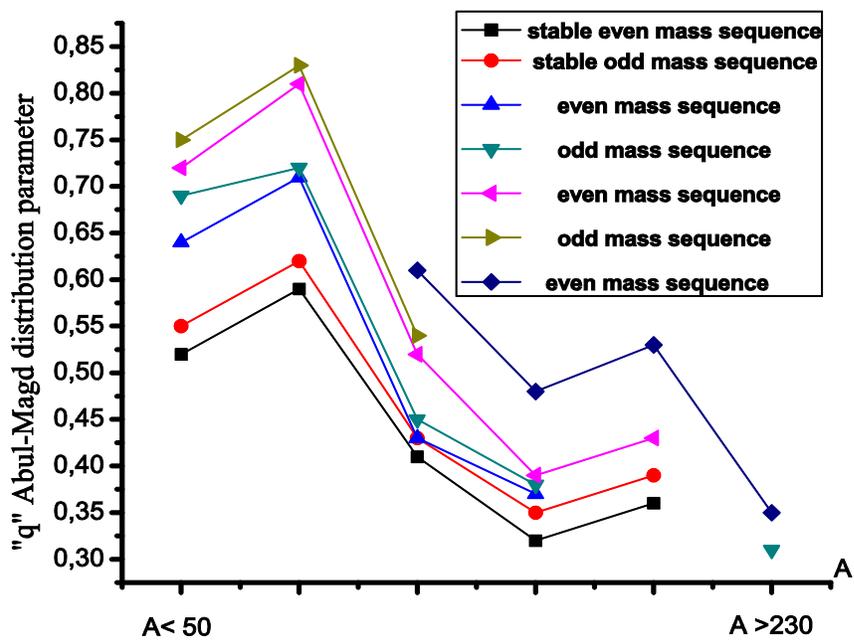

Figure3.

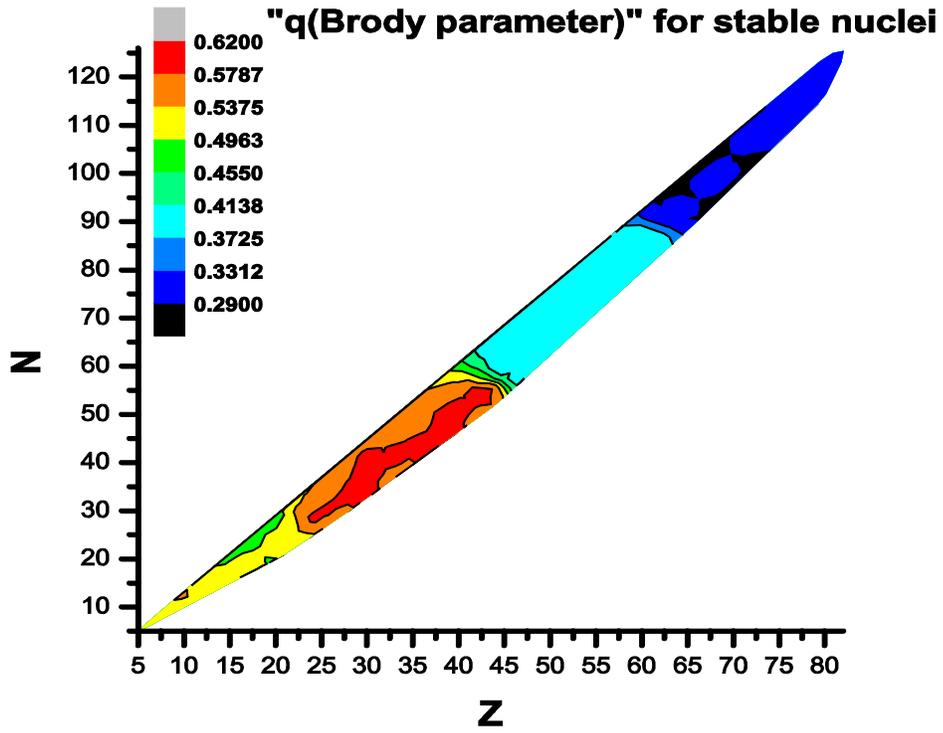

Figure4.

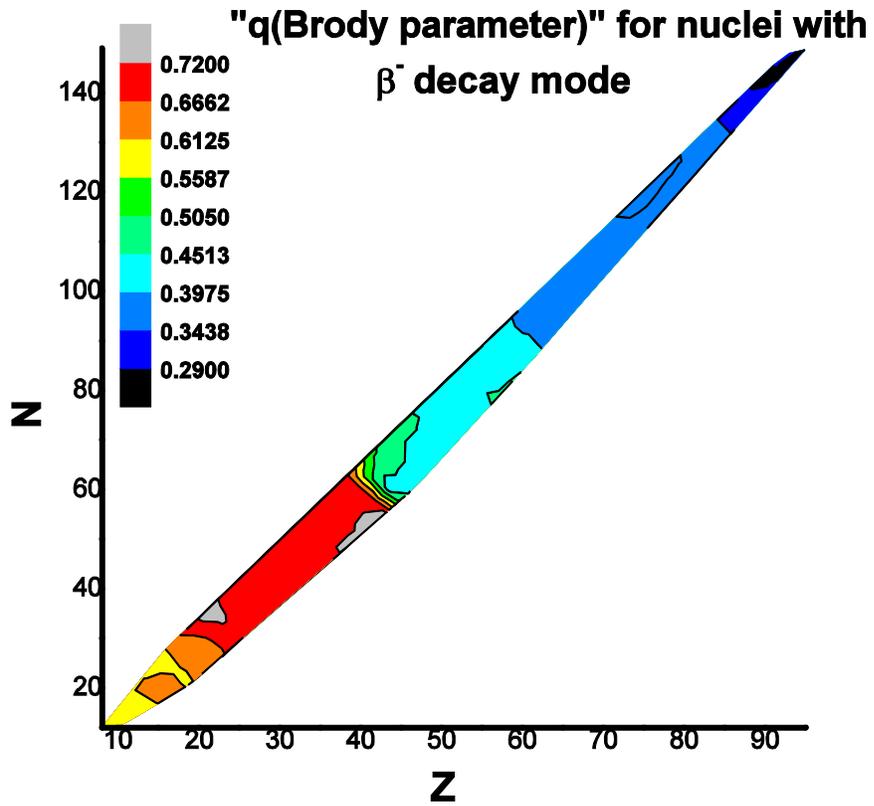

Figure5.

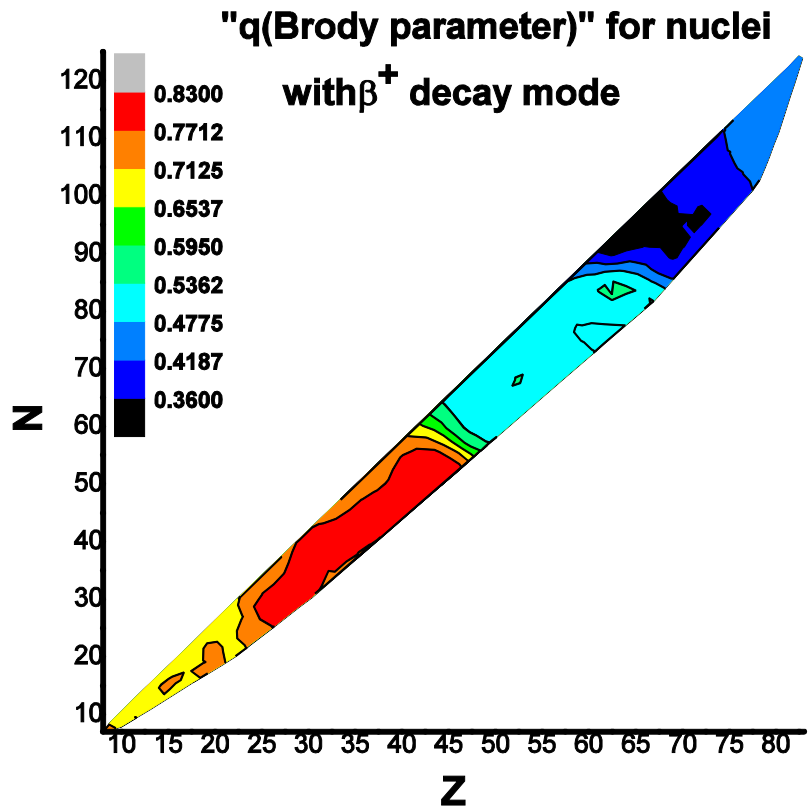

Figure6.

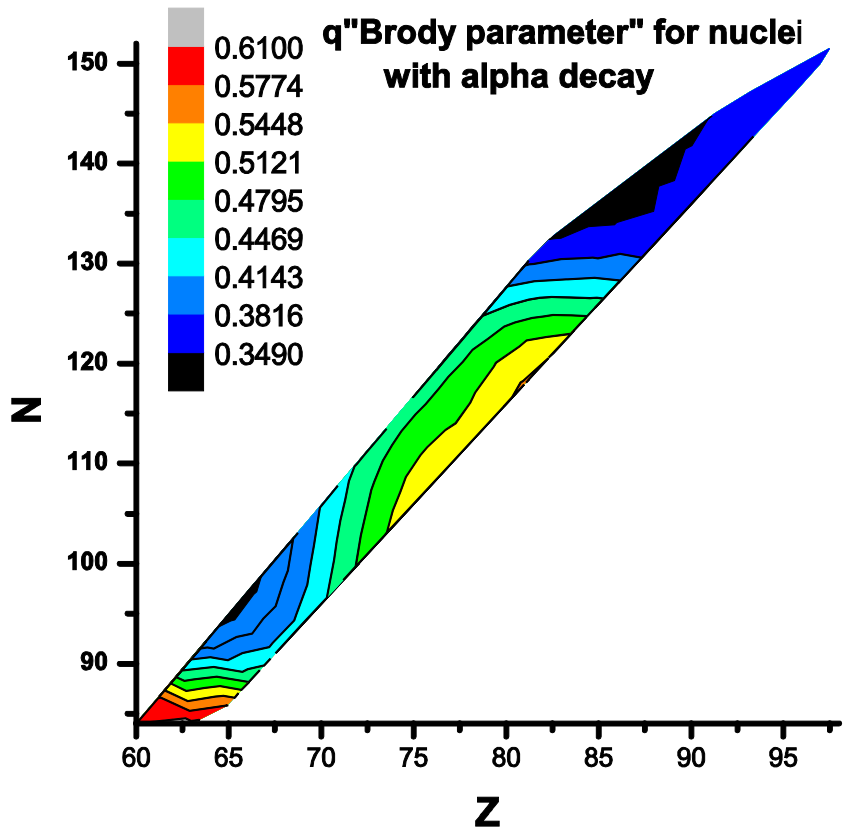

Figure7.

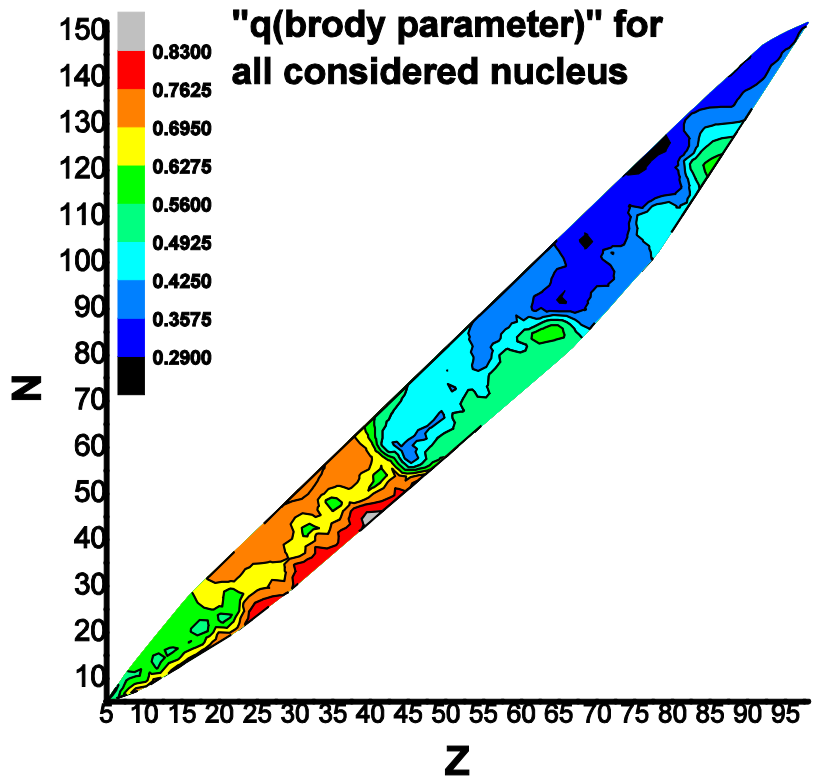

Figure8.

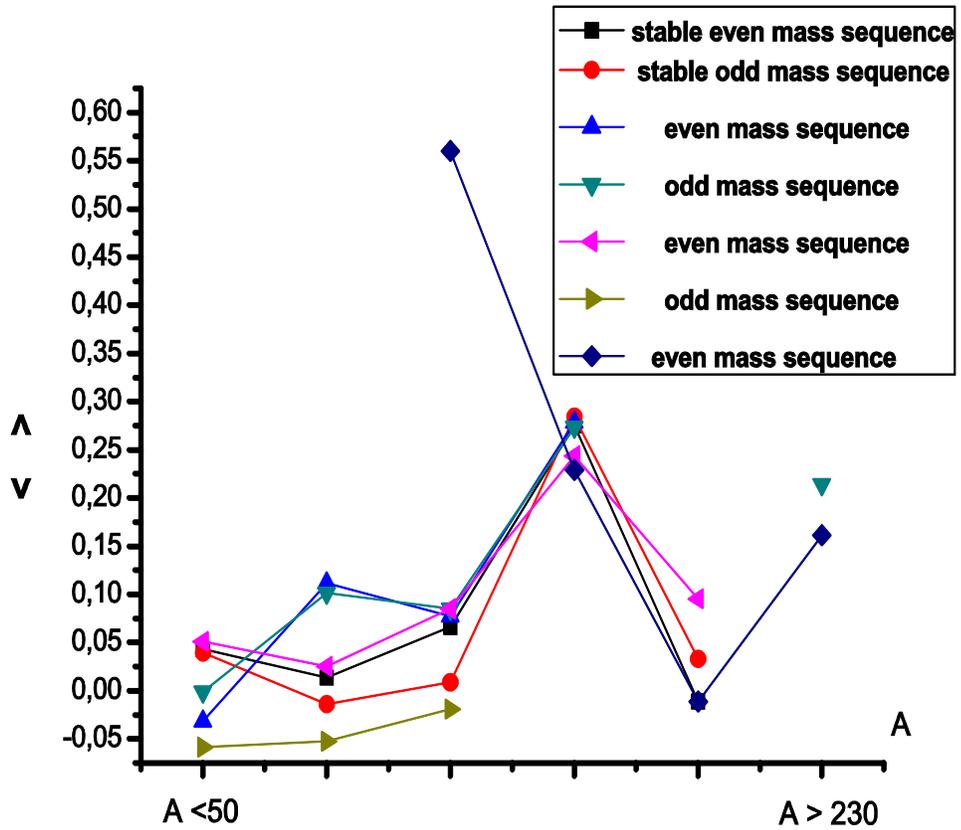